\documentclass[10pt,twocolumn,letterpaper]{article}

\usepackage{iccv}
\usepackage{times}
\usepackage{epsfig}
\usepackage{graphicx}
\usepackage{amsmath}
\usepackage{amssymb}

\usepackage[breaklinks=true,bookmarks=false]{hyperref}
\usepackage[htt]{hyphenat}
\usepackage{xcolor}

\iccvfinalcopy 

\def\iccvPaperID{3727} 

\ificcvfinal\fi

\begin{document}

\title{\vspace{-0.5in} Recovering a Molecule's 3D Dynamics from Liquid-phase Electron Microscopy Movies \vspace{-0.1in}}

\author{
Enze Ye\textsuperscript{1,2,4}\;  
Yuhang Wang\textsuperscript{4}\;  
Hong Zhang\textsuperscript{3}\;  
Yiqin Gao\textsuperscript{3}\;  
Huan Wang\textsuperscript{1,3}\; 
He Sun\textsuperscript{1,2}\thanks{Corresponding author}\\ 
\textsuperscript{1}National Biomedical Imaging Center, Peking University, Beijing, China.\\
\textsuperscript{2}College of Future Technology, Peking University, Beijing, China.\\
\textsuperscript{3}College of Chemistry and Molecular Engineering, Peking University, Beijing, China.\\
\textsuperscript{4}DP Technology, Ltd., Beijing, China.\\
\texttt{\small yez23@stu.pku.edu.cn, wangyh@dp.tech, zhangh@pku.edu.cn,} \\
\texttt{\small gaoyq@pku.edu.cn, wanghuan\_ccme@pku.edu.cn, hesun@pku.edu.cn}
\vspace{-0.1in}}

\maketitle
\ificcvfinal\thispagestyle{empty}\fi

\begin{abstract}
The dynamics of biomolecules are crucial for our understanding of their functioning in living systems. However, current 3D imaging techniques, such as cryogenic electron microscopy (cryo-EM), require freezing the sample, which limits the observation of their conformational changes in real time. The innovative liquid-phase electron microscopy (liquid-phase EM) technique allows molecules to be placed in the native liquid environment, providing a unique opportunity to observe their dynamics. In this paper, we propose TEMPOR, a Temporal Electron MicroscoPy Object Reconstruction algorithm for liquid-phase EM that leverages an implicit neural representation (INR) and a dynamical variational auto-encoder (DVAE) to recover time series of molecular structures. We demonstrate its advantages in recovering different motion dynamics from two simulated datasets, 7bcq \& Cas9. To our knowledge, our work is the first attempt to directly recover 3D structures of a temporally-varying particle from liquid-phase EM movies. It provides a promising new approach for studying molecules' 3D dynamics in structural biology.
\end{abstract}

\section{Introduction} 
\label{sec:intro}
The study of the structures and dynamics of biological molecules is crucial for understanding their functions and interactions in living systems. Electron microscopy (EM) has become an indispensable tool for providing high-resolution images and 3D structures of biological molecules at the nanoscale level, without the need for crystallization. Cryogenic electron microscopy (cryo-EM)\cite{nogales2016development}, the Nobel-prize-winning innovation, has played a crucial role in this development. By rapidly freezing the purified sample solution and subsequently reconstructing a 3D scattering potential (i.e. volume) from $10^{4-7}$ 2D projection images taken from various angles, cryo-EM has enabled researchers to obtain precise structures of complex biomolecules, such as proteins, viruses and ribosomes. This has led to many breakthroughs in structural biology. However, cryo-EM requires the sample to be immobilized, which limits its capability to study dynamic processes, such as chemical reactions and conformational changes.

Recent advances in liquid-phase EM\cite{lyu2023electron, wang2021imaging, wu2020liquid} have opened up new possibilities for studying the dynamic behavior of biological molecules in real time. By placing samples in the liquid medium, liquid-phase EM allows molecules to move and interact in a natural environment, consequently capturing movies of functional variations in their structures. However, the current liquid-phase EM technique primarily focuses on observing the dynamics of molecules in two dimensions, rather than recovering their 3D structures. A typical liquid-phase EM dataset contains only $10^{3-4}$ noisy 2D images. Due to the limited number of particles that can be imaged and the temporal evolution of each particle, it is extremely challenging to extract 3D structural information from the liquid-phase EM data. 

\begin{figure}
\centering
\includegraphics[width=75mm]{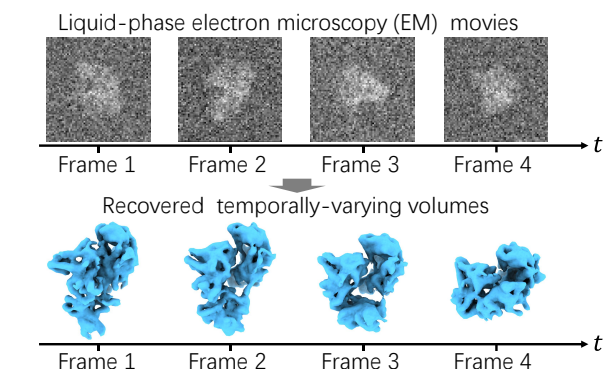}
\caption{Problem definition. Liquid-phase EM reconstruction aims to recover a 3D volume for each frame of its movies.}
\label{Figure0:task}
\end{figure}

In this paper, we propose a novel dynamic tomographic reconstruction algorithm for liquid-phase EM (Fig.~\ref{Figure0:task}), called \nohyphens{TEMPOR} (Temporal Electron MicroscoPy Object Reconstruction). The key insight of this algorithm is that despite liquid-phase EM collecting a much smaller number of images than cryo-EM, the temporal information in each particle's movie can provide additional supervision for the volume reconstruction. This algorithm builds upon recent advances in heterogeneous cryo-EM reconstruction\cite{zhong2021cryodrgnnm, zhong2019reconstructing, zhong2021cryodrgn2}, and leverages an implicit neural representation (INR) and a dynamical variational auto-encoder (DVAE) to recover the time series of a molecule's 3D structural variations. To our knowledge, our paper is the first work that tries to reconstruct 3D structures of temporally-varying particles directly from liquid-phase EM movies. This has important implications for various areas of research, including precision medicine, molecular biology, and biophysics. 
\section{Background and Problem Definition} 
\label{sec:background}
\subsection{Image Formation Model of EM} \label{subsec:formation}
The primary objective of both cryo-EM and liquid-phase EM is to reconstruct a molecule's 3D scattering potential (volume) from a collection of its 2D projection images. The formation model of an EM image can be represented as
\begin{equation} \label{eq:formation}
\begin{split}
    &X(r_x, r_y) = g * \int_{\mathbb{R}} V(R^{T}\mathbf{r}+\mathbf{t})dr_{z} + n,\\
    &\mathbf{r} = (r_x, r_y, r_z)^T, \ \mathbf{t}=(t_x, t_y, 0)^T
\end{split}
\end{equation}
where $X$ denotes the 2D EM image, $V$ denotes the molecule's 3D volume, $g$ is the microscope's point spread function caused by electron diffraction, $n$ is the measurement noise, and $R \in SO(3)$ and $\mathbf{t}$ are the unknown orientation and translation of the molecule, respectively.

By taking Fourier transforms ($\mathcal{F}_{2D}$ or $\mathcal{F}_{3D}$) of both sides of Eq.~\ref{eq:formation}, one can derive the image formation model in the frequency domain,
\begin{equation} \label{eq:fourierslice}
    \hat{X}(k_x, k_y)=\hat{g}S(\mathbf{t})\hat{V}(R^T(k_x, k_y, 0)^T) + \hat{n}
\end{equation}
where $\hat{X}=\mathcal{F}_{2D}\{X\}$ and $\hat{V}=\mathcal{F}_{3D}\{V\}$ represent the Fourier transformed image and volume, $\hat{g}=\mathcal{F}_{2D}\{g\}$ is the microscope's Contrast Transfer Function (CTF), $S(\mathbf{t})$ is the phase shift operator defined by the spatial domain translation, $\hat{n}=\mathcal{F}_{2D}\{n\}$ is the frequency-dependent measurement noise, and $(k_x, k_y)$ is the frequency coordinate. Equation~\ref{eq:fourierslice} defines a modified version of the \textit{Fourier slice theorem}, which states that the Fourier transformed 2D projection of a volume equals the central slice from the Fourier transformed 3D volume in the perpendicular direction. This relationship simplifies the connection between the 2D projection measurements and the underlying 3D volume. 

\subsection{Cryo-EM Reconstruction} \label{subsec:cryoEM}
Cryo-EM collects images of frozen single particles with unknown poses, denoted as $\mathcal{D}_{\text{cryo}}=\{X_i; i\in [1, N]\}$, to reconstruct a molecule's 3D volume(s). Depending on whether different particles share the same structure, we conduct either homogeneous or heterogeneous reconstruction of the molecule's 3D volume(s). 

Homogeneous reconstruction assumes the volumes of all particles are identical, so different EM images are simply projections of a fixed structure from different angles. The volume, $\hat{V}$, and the pose of each particle, $\phi_i = \{R_i, t_i\}$, can be jointly solved using a \textit{maximum a posteriori} (MAP) formulation,
\begin{equation} \label{eq:homogeneous}
\begin{split}
    \hat{V}^{\star}, \phi_i^{\star} = \underset{\hat{V}, \phi_i}{\arg \max}\{\sum_{i=1}^N \log p(\hat{X}_i|\phi_i, \hat{V}) \\ 
    + \sum_{i=1}^N \log p(\phi_i) + \log p(\hat{V})\}, 
\end{split}
\end{equation}
where $p(\hat{X}_i|\phi_i, \hat{V})$ is the probability of observing an image $\hat{X}_i$ with pose $\phi_i$ from volume $\hat{V}$, $p(\phi_i)$ is the prior distribution of a particle's pose, and $p(\hat{V})$ is the prior distribution of a particle's volume. Since the measurement noise typically satisfies an independent, zero-mean Gaussian distribution, it is straightforward to derive the probabilistic model, $p(\hat{X}_i|\phi_i, \hat{V})$, based on the image formation model in Eq.~\ref{eq:fourierslice}. 

Heterogeneous reconstruction relaxes the assumption that different particles share the same structure, allowing for the recovery of multiple independent volumes, $\{\hat{V}_j^{\star}; j \in [1, M]\}$, from a set of cryo-EM images. This involves optimizing the following expression,
\begin{equation} \label{eq:heterogeneous}
\begin{split}
    \hat{V}_j^{\star}, \phi_i^{\star}, \pi_{i, j}^{\star} = \underset{\hat{V}_j, \phi_i, \pi_{i, j}}{\arg \max} \{\sum_{i=1}^N \log \sum_{j=1}^M \pi_{i,j} p(\hat{X}_i|\phi_i, \hat{V}_j) \\
    + \sum_{i=1}^N \log p(\phi_i) + \sum_{j=1}^M \log p(\hat{V}_j)\},
\end{split}
\end{equation}
where $\pi_{i, j}$ is the probability of image $\hat{X}_i$ assigned to volume $\hat{V}_j$. In addition to recovering the volumes and poses, this optimization problem also needs to cluster particles into different volume categories. The number of volume categories $M$ is a user-defined hyper-parameter that measures structural heterogeneity in a cryo-EM dataset. The reconstruction problem becomes increasingly challenging as $M$ increases.

\subsection{Liquid-phase EM Reconstruction} \label{subsec:liquidEM}
Liquid-phase EM captures movies of freely-moving particles, denoted as $\mathcal{D}_{\text{liquid}} = \{X_{i, t}; i \in [1, N], t \in [1, T]\}$, to recover the dynamics of a molecule's 3D volume. A liquid-phase EM dataset typically contains tens to hundreds of movies, where each movie consists of 10-100 frames and represents a sequence of projection images from an evolving particle. Compared to cryo-EM, the total number of images in a liquid-phase EM dataset is two to three orders of magnitude smaller. Moreover, because a particle's volume varies over time, we need to compute a $\hat{V}_{i, t}$ for each frame of a liquid-phase EM movie $\hat{X}_{i, t}$. This gives rise to a highly ill-posed reconstruction inverse problem, which can be expressed as follows,
\begin{equation} \label{eq:liquid}
\begin{split}
    \hat{V}_{i, t}^{\star}, \phi_{i, t}^{\star} = &\underset{\hat{V}_{i, t}, \phi_{i, t}}{\arg \max} \{\sum_{i=1}^N \sum_{t=1}^T \log p(\hat{X}_{i, t}|\phi_{i, t}, \hat{V}_{i, t}) \\
    &+ \sum_{i=1}^N \sum_{t=1}^T \log p(\phi_{i, t}) + \sum_{i=1}^N \sum_{t=1}^T \log p(\hat{V}_{i, t})\}.
\end{split}
\end{equation}
If we ignore the temporal dependency in a movie, this problem becomes equivalent to a heterogeneous cryo-EM reconstruction task with the same number of volume categories and images. Due to its limited data and massive unknown variables, this problem remains an open challenge in liquid-phase EM, and consequently, current research using liquid-phase EM has rarely explored the 3D structures of a molecule. In the following sections, we will present our proposed solution to this problem and demonstrate its effectiveness through a series of numerical experiments.
\section{Related Work} \label{sec:related}
\subsection{Homogeneous \& Heterogeneous Cryo-EM Reconstruction} \label{subsec:recon}
A common approach for homogeneous cryo-EM reconstruction is the expectation-maximization algorithm\cite{scheres2012bayesian}, which alternates between solving for the unknown 3D volume or the particles' poses while fixing the other. This algorithm refines the 3D molecule structure iteratively until the image formation model predicts 2D images that match the EM measurements, hence it is also known as \textit{iterative refinement} in structural biology. A modified version of \textit{iterative refinement} algorithm, termed \textit{multi-class refinement}, can be extended to heterogeneous cryo-EM reconstruction when the number of volume categories is small. \textit{Multi-class refinement} algorithm loops between expert-guided clustering of cryo-EM images and homogeneous reconstruction of each independent group, finally providing a structure for each cluster of images. Many popular cryo-EM analysis tools, such as FREALIGN\cite{lyumkis2013likelihood}, CryoSPARC\cite{punjani2017cryosparc} and RELION\cite{nakane2018characterisation}, have been developed using this refinement framework. However, refinement-based approaches struggle to achieve robust reconstruction when the molecule's structural heterogeneity is complex. Recent heterogeneous reconstruction methods such as CryoDRGN\cite{zhong2021cryodrgnnm, zhong2019reconstructing, zhong2021cryodrgn2} have proposed modeling heterogeneity using a deep generative model with continuous latent states instead of discrete categories. This innovative formulation enables an infinite number of volume categories to be captured, allowing for the recovery of continuous conformational changes from the cryo-EM data. CryoFire\cite{levy2022amortized} and HetEM\cite{chen2023acehetem} further enhanced training speed and robustness by upgrading CryoDRGN's search-based pose estimation to an amortized inference neural network.

\subsection{Dynamic Computed Tomography} \label{subsec:dynamicCT}
Dynamic computed tomography (CT) is a challenging but essential problem in various fields, including security, industry, and healthcare. In dynamic CT, the object's temporal variation leads to the acquisition of only sparse angular views during CT scanning, making it difficult to reconstruct the object's structure at each moment. Classical dynamic CT techniques utilize parametric models to suppress periodic motions (e.g., cardiac or respiratory motions in clinical CT) for high-quality volume reconstruction\cite{rohkohl2008c, schwemmer2013residual, schwemmer2013opening}. Advanced dynamic CT methods can handle more complicated non-periodic motions by jointly solving a static scene and a motion field for the recovery of the object's dynamics\cite{mohan2015timbir, zang2018space, zang2019warp}. Recent research has proposed using novel deep learning techniques from 3D computer vision to improve the modeling of time-dependent scene and motion fields in dynamic CT\cite{reed2021dynamic}. However, most existing dynamic CT techniques require slow object motion and low-noise measurements to generate high-quality results. These limitations make them inapplicable for solving liquid-phase EM reconstruction, which defines a unique dynamic CT problem aiming to reconstruct a 3D volume from every single view despite having very noisy measurements.

\subsection{Implicit Neural Representation} \label{subsec:INR}
Implicit Neural Representation (INR) has proven to be a powerful technique in computer vision and graphics applications. This approach is based on the concept of ``coordinate-based" neural networks\cite{mescheder2019occupancy, mildenhall2021nerf, park2019deepsdf, sitzmann2020implicit} that learn functions mapping 3D coordinates to physical properties of the scene. INR's simple network structure enables it to be less prone to overfitting and utilize its inherent function smoothness to regularize unobserved views of scenes. It has achieved remarkable success in graphical rendering of static and dynamic scenes from limited views\cite{gafni2021dynamic, li2021neural, park2021nerfies, peng2021animatable, peng2021neural, tretschk2021non}, as well as in scientific imaging applications, including clinical X-ray CT\cite{corona2022mednerf}, surgical endoscopy\cite{wang2022neural}, optical microscopy\cite{cao2022dynamic, liu2022recovery}, and black hole emission tomography\cite{levis2022gravitationally}. The latest cryo-EM reconstruction algorithms, such as CryoDRGN\cite{zhong2021cryodrgnnm, zhong2019reconstructing, zhong2021cryodrgn2} and Cryo-AI\cite{levy2022cryoai}, also use INR to model the Fourier domain of a molecule's 3D structure. 


\subsection{Dynamical Variatioinal Auto-Encoder} \label{subsec:DVAE}
Dynamical variational auto-encoder (DVAE)\cite{girin2020dynamical} is a versatile class of deep generative models that has a wide range of applications in spatial-temporal analysis. This type of model inherits the fundamental characteristics of variational auto-encoder (VAE)\cite{kingma2013auto}, which transforms high-dimensional data into a low-dimensional latent space. Additionally, it incorporates deep temporal inference models, such as recurrent neural networks and state space models, to capture the temporal dependence of the learned latent states. DVAE has found numerous applications in vision tasks, including human motion estimation\cite{chen2021dynanet} and video generation\cite{lombardo2019deep, yingzhen2018disentangled}, and has also demonstrated remarkable performance in scientific fields such as dynamical system identification in physics\cite{lopez2020variational, takeishi2017learning} and health monitoring in biomedicine\cite{krishnan2015deep}. By extracting low-dimensional dynamics from spatial-temporal data in an unsupervised manner, DVAE offers a natural way to analyze the temporal variation of complex systems. In the following context, we will utilize a specific type of DVAE called structured inference network\cite{krishnan2017structured, krishnan2015deep} to model the molecule's temporal variation in a movie.

\section{Methods} \label{sec:methods}
\begin{figure}
\centering
\includegraphics[width=80mm]{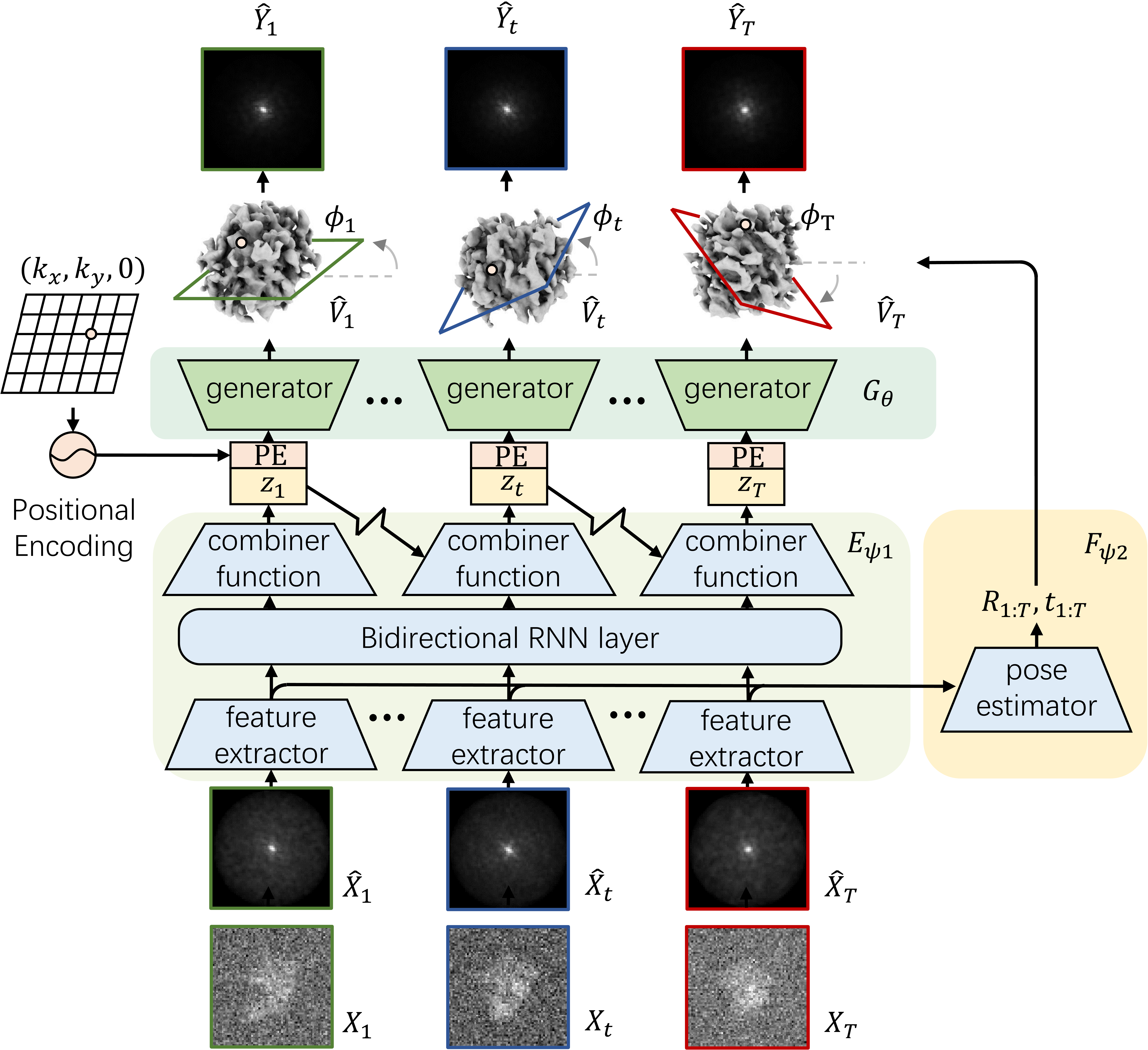}
\caption{Overview of the TEMPOR method. This method utilizes a dynamic VAE to reconstruct real-time 3D dynamics of molecules from liquid-phase EM movies. First, movie ($X_{1:T}$) is transformed into the Fourier domain ($\hat{X}_{1:T}$) and passed through the encoder, a temporal inference model ($E_{\psi_1}$), to infer the molecule's latent states ($z_{1:T}$). At each moment the latent state ($z_{t}$) depends on the input movie and the previous latent state ($z_{t-1}$), and thus captures the low-dimensional dynamics of a molecule's conformational changes. These latent states are then used by the INR-based volume generator ($G_{\theta}$) to create temporally-varying volumes ($\hat{V}_t$) and corresponding projection images ($\hat{Y}_t$) from views ($\phi_t = \{R_t, t_t\}$) estimated by pose estimator ($F_{\psi_2}$). The parameters $\psi=\{\psi_1, \psi_2\}$ and $\theta$ are jointly optimized by minimizing the difference between input and output images, resulting in a reconstructed time series of a molecule's 3D structures.}
\label{Figure1:overview}
\end{figure}

In this section, we introduce our method, TEMPOR (Temporal Electron MicroscoPy Obejct Reconstruction), which combines INR and DVAE to recover 3D dynamics of molecules from liquid-phase EM movies. We first present the probabilistic formulation of this algorithm in Sec.~\ref{subsec:algorithm}. Then, we describe the implementation details on neural network architectures and training strategies in Sec.~\ref{subsec:implementation}.

\subsection{Probabilistic Formulation of Liquid-phase EM Reconstruction} \label{subsec:algorithm}
Liquid-phase EM reconstruction can be formulated as an MAP estimation problem in Eq.~\ref{eq:liquid}, which aims to recover a 3D volume, $\hat{V}_{i, t}$, and its pose, $\phi_{i, t}$, for each frame of liquid-phase EM movies, $\mathcal{D}_{\text{liquid}} = \{X_{i, t}; i \in [1, N], t \in [1, T]\}$. This reconstruction problem can be denoted as follows,
\begin{equation} \label{eq:liquid2}
\begin{split}
    \underset{\hat{V}_{i, t}, \phi_{i, t}}{\max} &\sum_{i=1}^N \sum_{t=1}^T \log p(\hat{X}_{i, t}| \hat{V}_{i, t}, \phi_{i, t}) \\
    s.t. \ &\hat{V}_{i, t} \sim p_{\hat{V}}(\cdot),
\end{split}
\end{equation}
where $p_{\hat{V}}(\cdot)$ is the prior distribution of a molecule's 3D volume. Given that a molecule's conformational changes can often be represented using low-dimensional features, it is possible to parameterize the 3D volume distribution using a deep generative model, $G_{\theta}(\cdot)$, with continuous latent states $z_{i, t}$,
\begin{equation} \label{eq:liquid3}
\begin{split}
    \underset{\theta, z_{i, t}, \phi_{i, t}}{\max} &\sum_{i=1}^N \sum_{t=1}^T \log p(\hat{X}_{i, t}| \hat{V}_{i, t}, \phi_{i, t}) \\
    s.t. \ &\hat{V}_{i, t} = G_{\theta}(z_{i, t}), \\
    & z_{i, t} \sim q(\cdot),
\end{split}
\end{equation}
where $q(\cdot)$ is the prior distribution of the latent states.

Assuming that the latent states and the poses can be computed from the EM images using an amortized inference manner, the reconstruction problem can be rewritten as
\begin{equation} \label{eq:liquid4}
\begin{split}
    \underset{\theta, \psi}{\max} &\sum_{i=1}^N \sum_{t=1}^T \log p(\hat{X}_{i, t}| \hat{V}_{i, t}, \phi_{i, t}) \\
    s.t. \ &\hat{V}_{i, t} = G_{\theta}(z_{i, t}), \\
    & z_{i, t}, \phi_{i, t} \sim q_{\psi}(\cdot| \hat{X}_{i, t})\\
    & z_{i, t} \sim \mathcal{N}(0, 1).
\end{split}
\end{equation}
This formulation is in agreement with the state-of-the-art deep learning cryo-EM reconstruction method, CryoDRGN, which employs a standard VAE framework to solve for the heterogeneous particle volumes. The objective function optimized by CryoDRGN is the evidence lower bound (ELBO) of the EM images,
\begin{equation} \label{eq:liquid5}
\begin{split}
\mathcal{L}(\hat{X}_{i, t};, \theta, \psi) &= \sum_{i=1}^N \sum_{t=1}^T \{ - D_{KL}[q_{\psi}(z_{i, t} | \hat{X}_{i, t}) \| \mathcal{N}(0, 1)]\\ 
&+\mathbb{E}_{q_{\psi}(z_{i, t}, \phi_{i, t} | \hat{X}_{i, t})}[\log p(\hat{X}_{i, t}| G_{\theta}(z_{i, t}, \phi_{i, t})]\}.
\end{split}
\end{equation}
In CryoDRGN, the generator, $G_\theta(\cdot)$, is modeled as an INR, which effectively regularizes the heterogeneous volumes in a given dataset to be fundamentally similar, despite low-dimensional conformational changes. Consequently, even though only a single view is observed for each volume, reasonable heterogeneous reconstruction can still be achieved. CryoDRGN provides a baseline approach for liquid-phase EM reconstruction. 

However, CryoDRGN neglects the crucial temporal information in liquid-phase EM movies. In contrast to cryo-EM reconstruction, liquid-phase EM data comprise movies of multiple particles. The temporal dependency among these movie frames could provide additional regularization to enhance the reconstruction quality. Given that the number of images in a liquid-phase EM dataset is considerably smaller (two to three orders of magnitude) than a cryo-EM dataset, it becomes extremely important to leverage these temporal information in the reconstruction. Motivated by the structured inference network\cite{krishnan2017structured, krishnan2015deep}, we reformulate the representation of the latent state distribution as a temporal inference model $E_{\psi_1}$. Meanwhile, given the molecule's Brownian motion has a characteristic time notably shorter than its deformation, the pose estimator is still represented as a time-independent inference model $F_{\psi_2}$. Now the liquid-phase EM reconstruction problem can be written as,
\begin{equation} \label{eq:liquid6}
\begin{split}
    \underset{\theta, \psi}{\max} &\sum_{i=1}^N \sum_{t=1}^T \log p(\hat{X}_{i, t}| \hat{V}_{i, t}, \phi_{i, t}) \\
    s.t. \ &\hat{V}_{i, t} = G_{\theta}(z_{i, t}), \\
    & z_{i, t} \sim E_{\psi_1}(\cdot | z_{i, t-1}, \hat{X}_{i, 1:T}), \\
    & \phi_{i, t} \sim F_{\psi_2}(\cdot | \hat{X}_{i, t}), \\
    & z_{i, t} \sim \mathcal{N}(0, 1), \psi = \{\psi_1, \psi_2\}
\end{split}
\end{equation}
where each latent state, $z_{i, t}$, depends on its past state, $z_{i, t-1}$, as well as all frames in a particle's movie, $\hat{X}_{i, 1:T}$, but each pose, $\phi_{i, t}$, only depends on a single frame. The objective function of our TEMPOR algorithm for liquid-phase EM reconstruction now becomes 
\begin{equation} \label{eq:liquid7}
\begin{split}
&\mathcal{L}(\hat{X}_{i, t};, \theta, \psi) = \\
&\sum_{i=1}^N \sum_{t=2}^T \{-D_{KL}[E_{\psi}(z_{i, t} | z_{i, t-1}, \hat{X}_{i, 1:T}) \| \mathcal{N}(0, 1)] +\\
& \mathbb{E}_{E_{\psi_1}(z_{i, t} | z_{i, t-1}, \hat{X}_{i, 1:T}) F_{\psi_2}(\phi_{i, t} | \hat{X}_{i, t})}[\log p(\hat{X}_{i, t}| G_{\theta}(z_{i, t}, \phi_{i, t}))] \}.
\end{split}
\end{equation}

\subsection{Implementation Details} \label{subsec:implementation}
Based on the probabilistic model in Eq.~\ref{eq:liquid6}, the neural network architecture for our TEMPOR algorithm is designed as depicted in Fig.~\ref{Figure1:overview}.

The volume generator $G_{\theta}(\cdot)$, inspired by CryoDRGN, is modeled as an INR in Fourier domain, which is a multi-layer perceptron (MLP) that takes latent state, $z_{i, t}$, and frequency coordinates $\mathbf{k} = (k_x, k_y, k_z)$ as its inputs and outputs the corresponding frequency components of a volume. Defining the volume generator in Fourier domain significantly improves training efficiency, as we only need to evaluate a 2D slice of coordinates for each movie frame, as suggested by the \textit{Fourier slice theorem}. Similar to other INR frameworks, the coordinates first pass a positional encoding block before being sent to the generator,
\begin{equation}
    \gamma(\mathbf{k}) = [\sin(\mathbf{k}), \cos(\mathbf{k}), \cdots, \sin(2^{L-1}\mathbf{k}), \cos(2^{L-1}\mathbf{k})],
\end{equation}
where $L$ is the maximum positional encoding degree. 

The temporal inference encoder, $E_{\psi_1}(\cdot)$, is modeled as a compound architecture that comprises a feature extractor, a bidirectional recurrent neural network (RNN) layer and a combiner function. The bidirectional RNN enables message passing along movie frames, allowing the latent state at each moment to depend on both the past state and the entire sequence of a particle's movie. For simplicity of computation, we assume that the latent state inferred by the encoder follows a mean field Gaussian distribution. The encoder's output is split in half, with the first half defining the mean of our latent state and the second half defining the diagonal components of its covariance matrix, 
\begin{equation}
E_{\psi_1}(\cdot) = \mathcal{N}(E_{\psi_1}^1(z_{i, t-1}, \hat{X}_{i, 1:T}), \text{diag}\{E_{\psi_1}^2(z_{i, t-1}, \hat{X}_{i, 1:T})\}).
\end{equation}

The pose estimation encoder, $F_{\psi_2}(\cdot)$, estimates the pose of each frame independently using a similar mean field Gaussian approach,
\begin{equation}
F_{\psi_2}(\cdot) = \mathcal{N}(F_{\psi_1}^1(\hat{X}_{i, t}), \text{diag}\{F_{\psi_1}^2(\hat{X}_{i, t})\}).
\end{equation}
As illustrated in Fig.~\ref{Figure1:overview}, $E_{\psi_1}(\cdot)$ and $F_{\psi_2}(\cdot)$ share the same feature extractor.

CryoDRGN and HetEM are used as baseline approaches in the numerical experiments to evaluate our method's performance. To ensure a fair comparison, our baseline implementation uses many identical building blocks as our TEMPOR algorithm, including an identical volume generator and an identical feature extractor in its latent encoder.

Due to the complex optimization problem defined by our method, we have observed that training our reconstruction network from scratch (ab initio) sometimes fails to converge to the optimal solution.  To overcome this challenge, we adopt a pretrain-finetune strategy in which we first train a baseline network that shares multiple identical build blocks with our TEMPOR algorithm. We then initialize our TEMPOR network with baseline's weights and train it for an extended duration. Empirical evidence has shown that this training strategy leads to higher quality reconstructions. See supplementary material for more implementation details. 

\section{Results}

\begin{figure*}[ht]
\centering
\includegraphics[width=160mm]{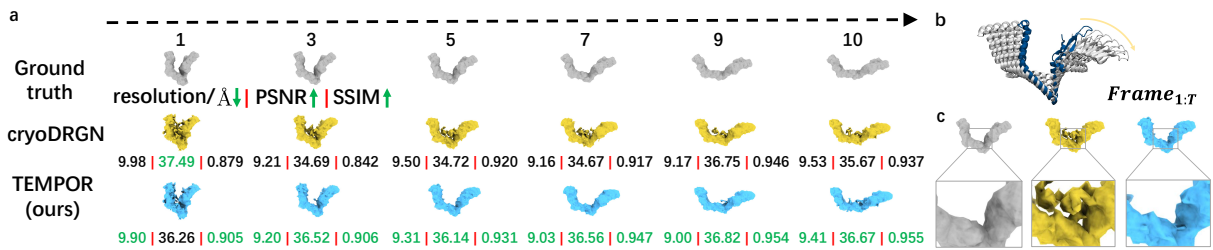}
\caption{Performance of TEMPOR on 7bcq dataset. \textbf{a.} Ground truth volumes and reconstructed volumes using TEMPOR and CryoDRGN for a randomly selected 7bcq movie (10 frames, frame 1, 3, 5, 7, 9, 10 displayed, SNR=0.1). TEMPOR successfully recovers the protein's linear rigid motion and generates volumes with better resolution, PSNR and SSIM values. 
\textbf{b.} One-dimensional peptide chain rotation of 7bcq protein. \textbf{c.} Detailed comparison of recovered volumes from frame 5.}
\label{Figure2:7bcq}
\end{figure*}

\begin{figure}
\centering
\includegraphics[width=80mm]{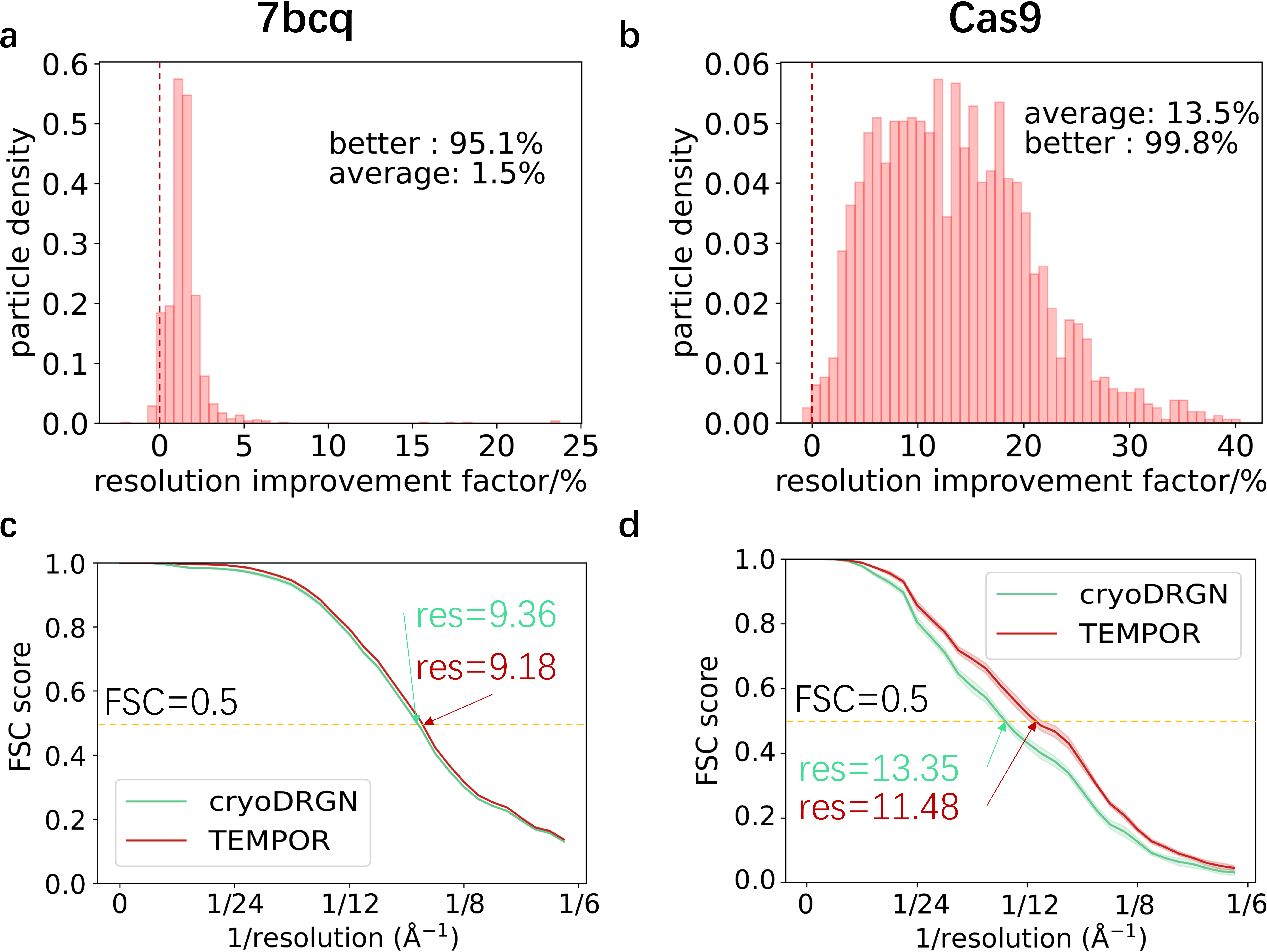}
\caption{\textbf{a, b.} Statistics of resolution improvement factors (ratios of TEMPOR's and CryoDRGN's resolutions minus 1) of all reconstructed volumes (100 movies $\times$ 10 frames in 7bcq and 100 movies $\times$ 19 frames in Cas9). 95.1\% and 99.8\% of TEMPOR's reconstructions outperform CryoDRGN's in 7bcq and Cas9 respectively, with an average improvement of 1.5\% and 13.5\%. \textbf{c, d.} Mean FSC curves (solid line) and their uncertainties (shadow) of all movies. TEMPOR achieves better resolution than CryoDRGN.}
\label{Figure4:analyze} 
\end{figure}

\begin{figure}
\centering
\includegraphics[width=80mm]{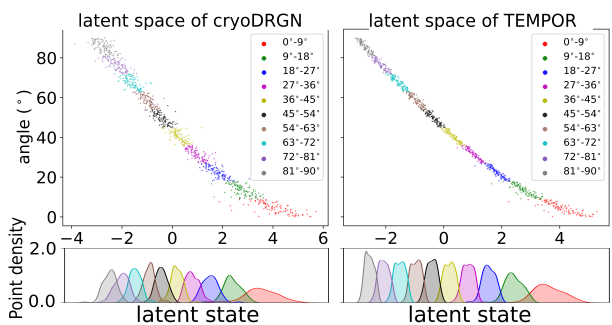}
\caption{(Top) Relationship between chain rotational angles and corresponding latent state values of all 7bcq movie frames. Different colors represent different volume categories classified based on ten rotational angle intervals. (Bottom) Latent state distribution of each category. Our proposed method, TEMPOR, exhibits a more compact latent space manifold than CryoDRGN, allowing for better clustering of different volume categories.}
\label{Figure4:7bcq_latent}
\end{figure}

In this section, we demonstrate the performance of our proposed method using two simulated liquid-phase EM datasets: 7bcq and Cas9. This section is structured as follows. In Sec.~\ref{subsec:datasets}, we first introduce the synthetic datasets and the evaluation metrics of our numerical experiments. Then, we present both qualitative and quantitative results of our method on the two datasets and compare their performance with CryoDRGN and HetEM. Specifically, in Sec.~\ref{subsec:7bcq} and Sec.~\ref{subsec:Cas9}, we compare CryoDRGN and our approach in a simplified setting under the assumption of known poses. Meanwhile, in Sec.~\ref{subsec:Cas9_hetem}, we evaluate the performance between HetEM and our method during simultaneous estimation of poses and volumes. Finally in Sec.~\ref{subsec:discussion}, we investigate the robustness of our approach under varying levels of noise and conduct a series of ablation studies on network architectures, training strategies, and other relevant factors. 

\begin{figure*}[ht]
\centering
\includegraphics[width=160mm]{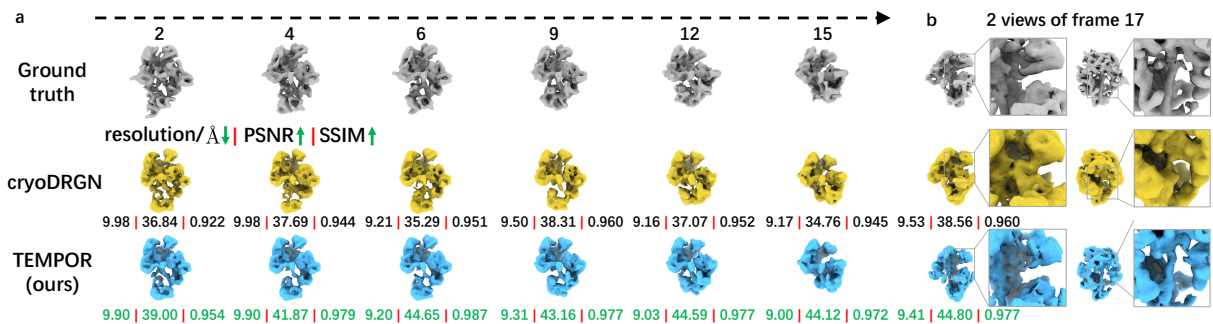}
\caption{\textbf{a.} Ground truth volumes and reconstructed volumes using TEMPOR and CryoDRGN for a randomly selected Cas9 movie (19 frames, frame 2, 4, 6, 9, 12, 15, 17 displayed, SNR=0.3). TEMPOR successfully captures the protein's deformational motion and generates valumes with better resolution, PSNR and SSIM values than CryoDRGN. \textbf{b.} Two views of the 17th frame are zoomed in to qualitatively compare the reconstruction details.}
\label{Figure3:cas9}
\end{figure*}

\begin{figure}
\centering
\includegraphics[width=60mm]{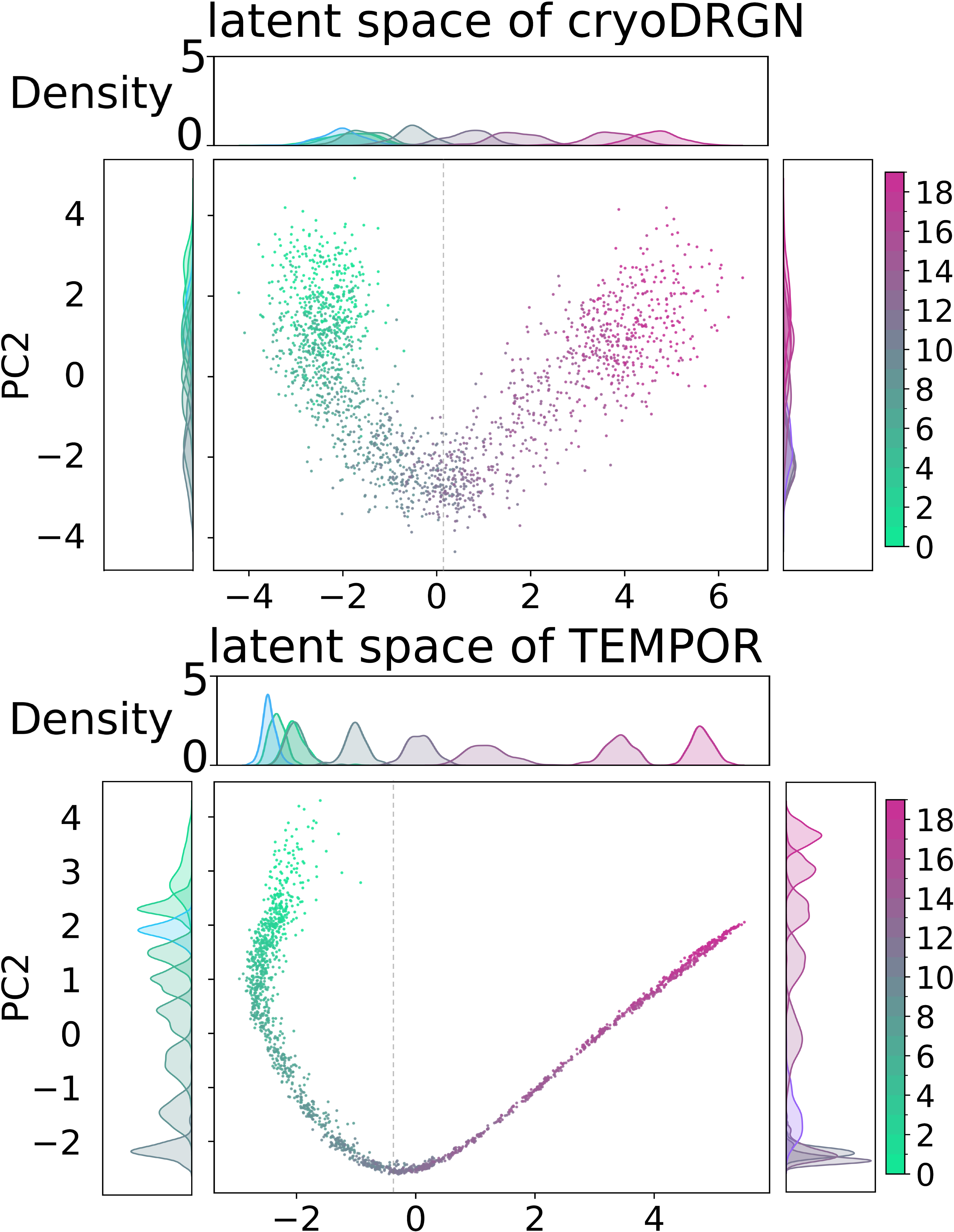}
\caption{Latent space manifolds of Cas9 dataset obtained through PCA. Each point represents a Cas9 movie frame, with different colors indicating different categories of Cas9's conformational changes. The distributions of each category's latent states are also projected onto the first two principal components (PCs), visualized on the top and sides of the figures. The distribution on the second PC is split into two parts to avoid overlapping. Our method, \nohyphens{TEMPOR}, exhibits a more compact latent space manifold than CryoDRGN, and better clusters different volume categories. }
\label{Figure4:cas9_latent}
\end{figure}

\subsection{Dataset \& Evaluation Metrics} \label{subsec:datasets}
Two simulated datasets of different levels of complexity were created for this study. The first dataset, 7bcq, is a simulated protein that undergoes motion primarily in one dimension. A short peptide chain of 7bcq rotates around a chemical bond at a constant speed (9$^\circ$ per frame) within a range of 90 degrees (Fig.~\ref{Figure2:7bcq}b). Our 7bcq dataset includes movies of 100 particles, each with 10 continuous frames. 

Cas9 is a more complex protein that undergoes non-rigid deformational motion. To generate multiple conformational changes of Cas9, we run molecular dynamics simulations that are initialized with two Cas9 structures from the protein data bank - the free (PDB ID 4CMQ) and RNA-bound (PDB ID 4ZT0) forms. We classify the resulting Cas9 volumes into 19 continuous states and created 500 synthetic movies, with each movie sequentially sampling a conformation from each of the 19 volume categories. In summary, the 7bcq dataset models simple one-dimensional rigid-body motions and contains 100 movies (10 frames each), while the Cas9 dataset models complex flexible structural deformations and contains 500 movies (19 frames each). See supplementary material for more details about the datasets.

Our reconstruction methods are evaluated using two major metrics: the manifold of learned latent space and the \textit{Fourier shell correlation} (FSC) curves of reconstructed volumes. The latent space manifold serves as a qualitative metric for evaluating the reconstruction performance, as a good reconstruction algorithm typically results in a latent space that well clusters different categories of volumes. On the other hand, the FSC coefficient is a widely accepted quantitative metric in electron microscopy reconstruction\cite{van2005fourier}. It measures the normalized cross-correlation between the reconstructed and true volumes, $\hat{V}_1$ and $\hat{V}_2$, as a function of radial shells in Fourier space (frequency radius $k$),
\begin{equation}
FSC(k) = \frac{\sum_{|\mathbf{k_i}| = k} \hat{V}_1(\mathbf{k_i}) \cdot \hat{V}_2(\mathbf{k_i})^{\star}}{\sqrt{\sum_{|\mathbf{k_i}| = k} |\hat{V}_1(\mathbf{k_i})|^2 \sum_{|\mathbf{k_i}| = k} |\hat{V}_2(\mathbf{k_i})|^2}}.
\end{equation}
Using FSC, we can also calculate the reconstruction resolution, which is the reciprocal of the maximum frequency radius, $k_{max}$, that yields a cross-correlation coefficient greater than 0.5. FSC and the resulting resolution provide a reliable measure of different algorithms' performance. For reference, we also report two standard computational imaging metrics of our reconstructed volumes: the Structural Similarity Index Measure (SSIM) and the Peak Signal-to-Noise Ratio (PSNR).

\begin{figure*}[ht]
\centering
\includegraphics[width=160mm]{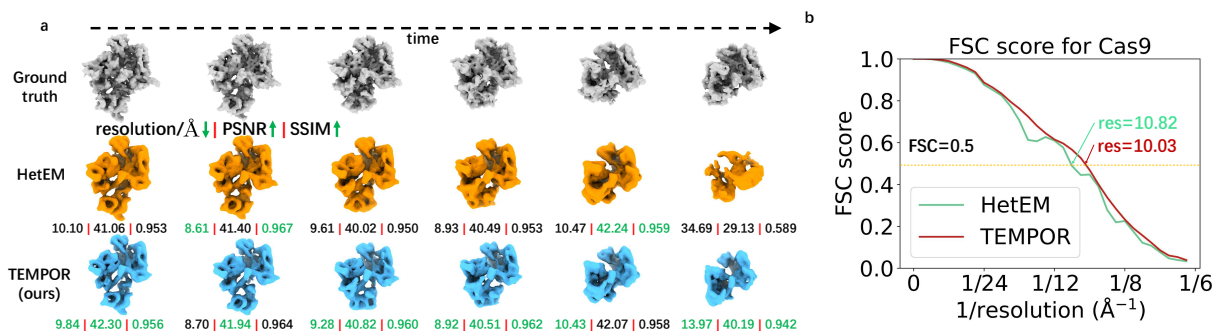}
\caption{\textbf{a.} Ground truth volumes and reconstructed volumes using HetEM and pose-estimation enabled TEMPOR for a randomly selected Cas9 movie (19 frames, frame 3, 6, 9, 12, 15, 18 displayed, SNR=1.0, 500 movies). TEMPOR outperforms the baseline method, HetEM, in most cases.  \textbf{b.} Mean FSC curves of all movies. TEMPOR achieves better resolution than HetEM, averaging $0.8\AA$ higher. }
\label{Figure5:abinit}
\end{figure*}

\subsection{7bcq: Reconstruction of Linear Rigid Motion} \label{subsec:7bcq}
In this section, we evaluate the performance of our method on recovering a linear rigid motion using the 7bcq dataset. As 7bcq's peptide chain rotation is a one-dimensional motion (Fig.~\ref{Figure2:7bcq}b), we set the latent state dimension of both CryoDRGN and our method to one.

Figure~\ref{Figure2:7bcq}a displays the reconstructed volumes obtained by our method and CryoDRGN, along with the ground truth 3D structures. This experiment was performed on a dataset with an SNR of 0.1, and the motion shown in Fig.~\ref{Figure2:7bcq}a was recovered from a randomly selected 7bcq movie with frames 1, 3, 5, 7, 9, 10 displayed. These results clearly demonstrate that our method produces higher-resolution 3D volumes with fewer errors compared to CryoDRGN, which sometimes exhibits unexpected spikes due to measurement noise. Additionally, we perform statistical analysis by computing the resolution improvement factor ($f_{i,t}$) for each movie frame
\begin{equation}
f_{i, t} = \frac{res_{i, t}^{cryoDRGN} - res_{i, t}^{ours}}{res_{i, t}^{cryoDRGN}},
\end{equation}
where $i$ and $t$ are movie and frame indices, and $res_{i, t}^{cryoDRGN}$ and $res_{i, t}^{ours}$ are the reconstruction resolutions obtained by CryoDRGN and our method, respectively. Our method reconstructs a higher resolution volume in 95.1\% of movie frames, and on average improves the resolution by 1.5\% compared to CryoDRGN (Fig.~\ref{Figure4:analyze}a). The mean FSC curve of all movies (Fig.~\ref{Figure4:analyze}c) indicates that, on average, our method increases the resolution by 0.18 \AA. We also evaluate PSNR and SSIM values for volumes recovered by CryoDRGN and TEMPOR after passing a low-pass filter, where TEMPOR also performs better in nearly all the cases.

The peptide-chain angles in the 7bcq movie are uniformly distributed between 0-90 degrees. In Fig.~\ref{Figure4:7bcq_latent}, we show the relationship between the true rotational angles and the latent state values estimated by CryoDRGN and our method, respectively. Our method demonstrates a more compact latent space manifold than CryoDRGN, suggesting better discrimination of images projected from volumes with small conformational differences. We further visualize the latent state distribution of movie frames from 10 different intervals of rotational angles ($0-9, 9-18, \cdots, 81-90$). Our method again reveals a stronger capability to distinguish small angle changes, suggesting its potential to improve the temporal resolution in liquid-phase EM reconstruction.

\begin{figure*}[ht]
\centering
\includegraphics[width=160mm]{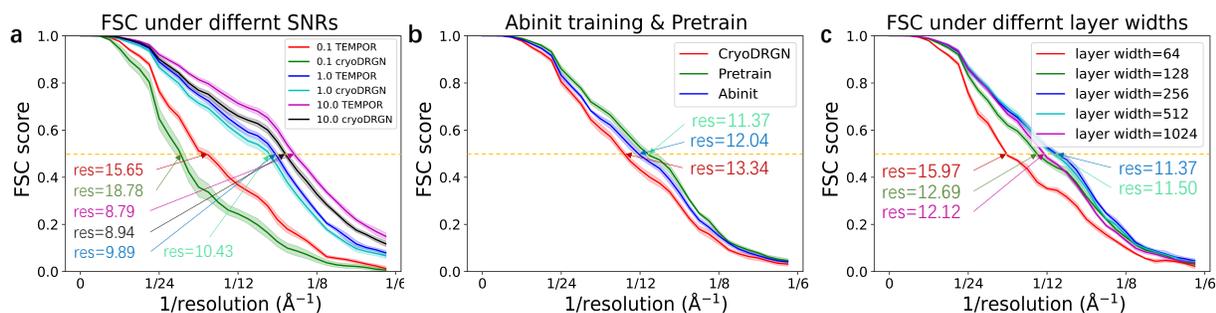}
\caption{\textbf{a.} Comparison of mean FSC curves (with uncertainty displayed as shadow) of TEMPOR and CryoDRGN under different SNRs (0.1, 1, 10). \textbf{b.} Mean FSC curves of CryoDRGN and TEMPOR using different training strategies. \textbf{c.} Mean FSC curves of TEMPOR using INR decoders with different layer widths. All experiments in these figures use the Cas9 dataset. \textbf{b.} and \textbf{c.} assume SNR=0.3.}
\label{fig:fsc_snr}
\end{figure*}

\begin{figure}[ht]
\centering
\includegraphics[width=80mm]{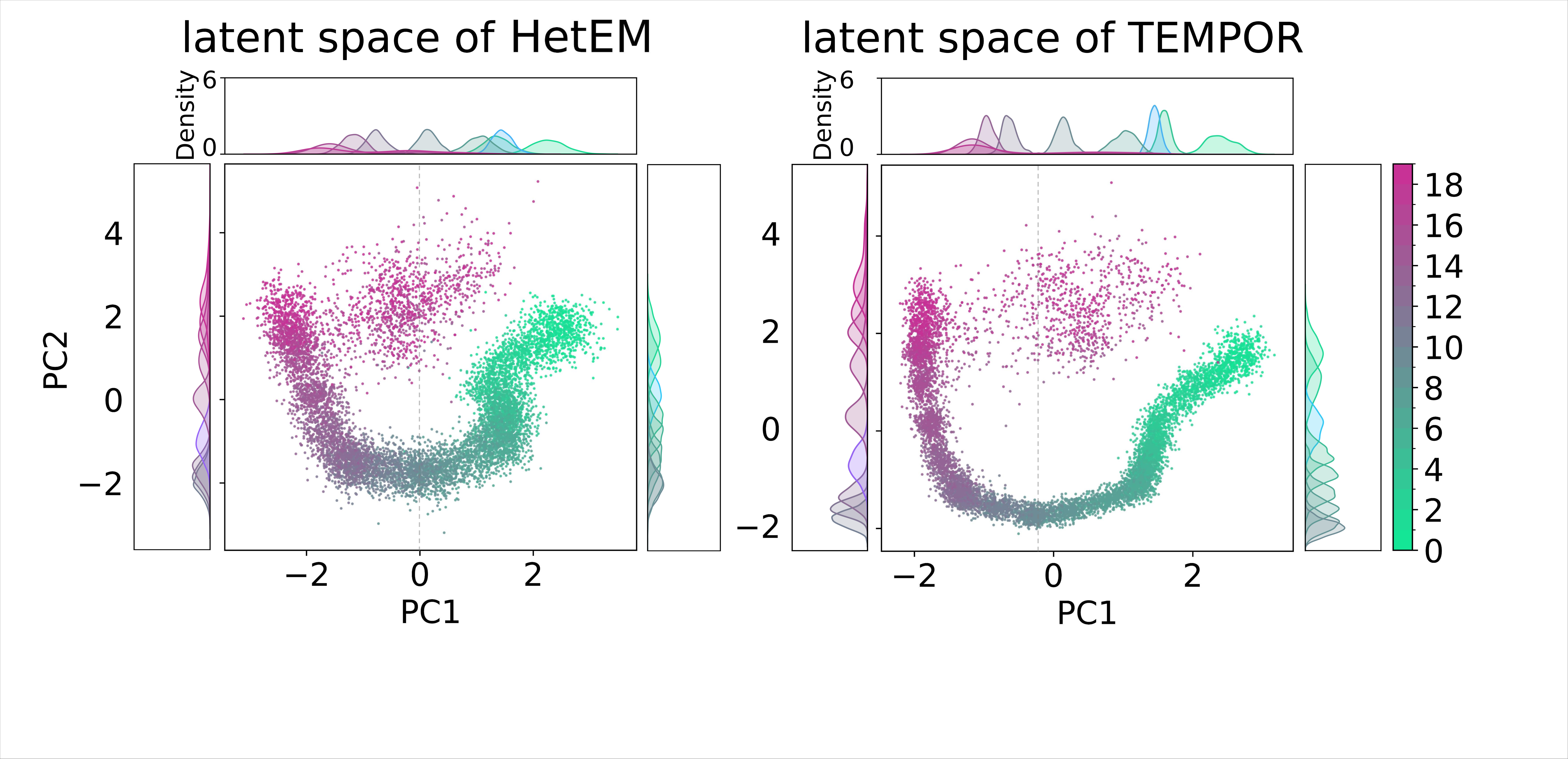}
\caption{Latent space manifolds of Cas9 dataset obtained through PCA (same as Fig.~\ref{Figure4:cas9_latent}). TEMPOR with pose estimation can better cluster different volume categories compared with HetEM.}
\label{Figure5:abinit_latent}
\end{figure}

\subsection{Cas9: Reconstruction of Deformational Motion}\label{subsec:Cas9}
In this section, we evaluate the performance of our method in recovering a more complex deformational motion using the Cas9 dataset. Biologically, this motion represents the protein's transformation from a free state to an RNA-bound state, which is more commonly observed in real experiments. To capture the complex conformational changes, we set the latent state dimensions to eight for both our method and CryoDRGN in all Cas9 experiments. Given that known poses simplify volume reconstruction, we use only 100 videos for training in this section.

Figure.~\ref{Figure3:cas9}a compares the 3D reconstruction of a randomly sampled Cas9 movie using CryoDRGN and our method, along with the ground-truth volumes. This experiment was performed on a dataset with an SNR of 0.3. Our method recovers finer structural details, such as the gap between protein domains and the central backbone in the 17th frame shown in Fig.~\ref{Figure3:cas9}b. Statistical metrics in Fig.~\ref{Figure4:analyze}b demonstrate the great advantages of our method in this example with complex deformational motions. It achieves better performance in 99.8\% of movie frames and on average improves the reconstruction resolution by 13.5\% compared to CryoDRGN. In Fig.~\ref{Figure4:analyze}d the mean FSC curve suggests that the resolution of our method outperforms CryoDRGN's by around 2 \AA. As shown in Fig.~\ref{Figure3:cas9}a, we also find that both PSNR and SSIM have greatly improved by TEMPOR.

In Fig.~\ref{Figure4:cas9_latent}, we visualize the 2-dimensional latent space manifolds obtained through PCA, with different colors representing the 19 conformational states in the Cas9 dataset. Our method shows a more compact manifold than CryoDRGN. As illustrated by both the 2D latent distribution and their 1D projections on each principal component, our method also displays better clustering ability.

\subsection{Joint estimation of poses and volumes}\label{subsec:Cas9_hetem}
In this section, we demonstrate the performance of our method when jointly estimating pose (rotation and translation) parameters and recovering 3D volumes. This experiment is conducted on the Cas9 dataset with SNR=1.0, 500 movies, and we compare the performance of TEMPOR with HetEM, a novel algorithm that considers the pose estimation but disregards temporal information. 

As illustrated in Fig.~\ref{Figure5:abinit}a and Fig.~\ref{Figure5:abinit_latent}, our method surpasses HetEM in both reconstruction quality and the compactness of latent state manifolds. Its reconstruction resolution outperforms that of HetEM by around $0.8\AA$ as shown in Fig.~\ref{Figure5:abinit}b. TEMPOR also generated volumes with higher PSNR and SSIM values for most projections.

\subsection{Discussion and Ablation Study} \label{subsec:discussion}
The previous numerical experiments on 7bcq and Cas9 have demonstrated the superior performance of our method in handling different motion dynamics. In this section, we investigate the robustness of our method under varying noise levels and report several ablation studies on neural network architectures and training strategies, all conducted on the Cas9 dataset.

In Fig.~\ref{fig:fsc_snr}a, we compare the FSC curves of CryoDRGN and our method under three SNR levels (0.1, 1, 10). While the two algorithms achieve comparable resolution at SNR=10, the gap between the two algorithms widens as SNR decreases, indicating the better robustness of our method in handling low-SNR liquid-phase EM data.

Figure~\ref{fig:fsc_snr}b compares the FSC curves of our method with and without the pre-training strategy proposed in Sec.~\ref{subsec:implementation}. Our method outperforms CryoDRGN even without pretraining, but a better training strategy further improves the reconstruction resolution by approximately 0.7 \AA. This justifies the necessity of advanced training techniques.

Figure~\ref{fig:fsc_snr}c compares the FSC curves obtained by varying neural network architectures. We experiment on INR decoders with different layer widths (number of neurons in each MLP layer). The results show that the reconstruction resolution first improves as the layer width increases, but then deteriorates when the layer width becomes too high. The layer width reflects the model capacity of an INR decoder. As the INR decoder models the prior distribution of the reconstructed volumes, this study suggests that the decoder architecture should not be overly simple or complex to achieve the best results. A careful design is necessary to balance the trade-off between capturing complex deformations and avoiding overfitting.

\section{Conclusion}
In this paper, we have proposed a novel deep learning tomographic reconstruction algorithm, called TEMPOR, which utilizes the temporal dependency of molecules' motions to recover their 3D dynamics from noisy liquid-phase EM movies. Our proposed method was evaluated in simulation on two datasets with different motion dynamics, 7bcq and Cas9, and outperformed the state-of-the-art methods, CryoDRGN and HetEM, in terms of both resolution and robustness to low-SNR data. It offers an initial conceptual illustration of recovering structures from temporally varying particles using EM observations.
\section{Acknowledgement}
This project has received funding from the National Natural Science Foundation of China(22174006). 
{\small
\bibliographystyle{ieee_fullname}
\bibliography{LEM_refs}

\begin{thebibliography}{10}\itemsep=-1pt

\bibitem{cao2022dynamic}
Ruiming Cao, Fanglin~Linda Liu, Li-Hao Yeh, and Laura Waller.
\newblock Dynamic structured illumination microscopy with a neural space-time
  model.
\newblock In {\em 2022 IEEE International Conference on Computational
  Photography (ICCP)}, pages 1--12. IEEE, 2022.

\bibitem{chen2021dynanet}
Changhao Chen, Chris~Xiaoxuan Lu, Bing Wang, Niki Trigoni, and Andrew Markham.
\newblock Dynanet: Neural kalman dynamical model for motion estimation and
  prediction.
\newblock {\em IEEE Transactions on Neural Networks and Learning Systems},
  32(12):5479--5491, 2021.

\bibitem{chen2023acehetem}
Weijie Chen, Lin Yao, Zeqing Xia, and Yuhang Wang.
\newblock {ACE-HetEM} for {\it{ab initio}} heterogenous cryo-{EM} {3D}
  reconstruction.
\newblock {\em arXiv:2308.04956}, 2023.

\bibitem{corona2022mednerf}
Abril Corona-Figueroa, Jonathan Frawley, Sam Bond-Taylor, Sarath Bethapudi,
  Hubert~PH Shum, and Chris~G Willcocks.
\newblock Mednerf: Medical neural radiance fields for reconstructing 3d-aware
  ct-projections from a single x-ray.
\newblock In {\em 2022 44th Annual International Conference of the IEEE
  Engineering in Medicine \& Biology Society (EMBC)}, pages 3843--3848. IEEE,
  2022.

\bibitem{gafni2021dynamic}
Guy Gafni, Justus Thies, Michael Zollhofer, and Matthias Nie{\ss}ner.
\newblock Dynamic neural radiance fields for monocular 4d facial avatar
  reconstruction.
\newblock In {\em Proceedings of the IEEE/CVF Conference on Computer Vision and
  Pattern Recognition}, pages 8649--8658, 2021.

\bibitem{girin2020dynamical}
Laurent Girin, Simon Leglaive, Xiaoyu Bie, Julien Diard, Thomas Hueber, and
  Xavier Alameda-Pineda.
\newblock Dynamical variational autoencoders: A comprehensive review.
\newblock {\em arXiv:2008.12595}, 2020.

\bibitem{kingma2013auto}
Diederik~P Kingma and Max Welling.
\newblock Auto-encoding variational bayes.
\newblock {\em arXiv preprint arXiv:1312.6114}, 2013.

\bibitem{krishnan2017structured}
Rahul Krishnan, Uri Shalit, and David Sontag.
\newblock Structured inference networks for nonlinear state space models.
\newblock In {\em Proceedings of the AAAI Conference on Artificial
  Intelligence}, volume~31, 2017.

\bibitem{krishnan2015deep}
Rahul~G Krishnan, Uri Shalit, and David Sontag.
\newblock Deep kalman filters.
\newblock {\em arXiv preprint arXiv:1511.05121}, 2015.

\bibitem{levis2022gravitationally}
Aviad Levis, Pratul~P Srinivasan, Andrew~A Chael, Ren Ng, and Katherine~L
  Bouman.
\newblock Gravitationally lensed black hole emission tomography.
\newblock In {\em Proceedings of the IEEE/CVF Conference on Computer Vision and
  Pattern Recognition}, pages 19841--19850, 2022.

\bibitem{levy2022cryoai}
Axel Levy, Fr{\'e}d{\'e}ric Poitevin, Julien Martel, Youssef Nashed, Ariana
  Peck, Nina Miolane, Daniel Ratner, Mike Dunne, and Gordon Wetzstein.
\newblock Cryoai: Amortized inference of poses for ab initio reconstruction of
  3d molecular volumes from real cryo-em images.
\newblock In {\em Computer Vision--ECCV 2022: 17th European Conference, Tel
  Aviv, Israel, October 23--27, 2022, Proceedings, Part XXI}, pages 540--557.
  Springer, 2022.

\bibitem{levy2022amortized}
Axel Levy, Gordon Wetzstein, Julien~NP Martel, Frederic Poitevin, and Ellen
  Zhong.
\newblock Amortized inference for heterogeneous reconstruction in cryo-em.
\newblock {\em Advances in Neural Information Processing Systems},
  35:13038--13049, 2022.

\bibitem{li2021neural}
Zhengqi Li, Simon Niklaus, Noah Snavely, and Oliver Wang.
\newblock Neural scene flow fields for space-time view synthesis of dynamic
  scenes.
\newblock In {\em Proceedings of the IEEE/CVF Conference on Computer Vision and
  Pattern Recognition}, pages 6498--6508, 2021.

\bibitem{liu2022recovery}
Renhao Liu, Yu Sun, Jiabei Zhu, Lei Tian, and Ulugbek~S Kamilov.
\newblock Recovery of continuous 3d refractive index maps from discrete
  intensity-only measurements using neural fields.
\newblock {\em Nature Machine Intelligence}, 4(9):781--791, 2022.

\bibitem{lombardo2019deep}
Salvator Lombardo, Jun Han, Christopher Schroers, and Stephan Mandt.
\newblock Deep generative video compression.
\newblock {\em Advances in Neural Information Processing Systems}, 32, 2019.

\bibitem{lopez2020variational}
Ryan Lopez and Paul~J Atzberger.
\newblock Variational autoencoders for learning nonlinear dynamics of physical
  systems.
\newblock {\em arXiv preprint arXiv:2012.03448}, 2020.

\bibitem{lyu2023electron}
Zhiheng Lyu, Lehan Yao, Wenxiang Chen, Falon~C Kalutantirige, and Qian Chen.
\newblock Electron microscopy studies of soft nanomaterials.
\newblock {\em Chemical Reviews}, 2023.

\bibitem{lyumkis2013likelihood}
Dmitry Lyumkis, Axel~F Brilot, Douglas~L Theobald, and Nikolaus Grigorieff.
\newblock Likelihood-based classification of cryo-em images using frealign.
\newblock {\em Journal of structural biology}, 183(3):377--388, 2013.

\bibitem{mescheder2019occupancy}
Lars Mescheder, Michael Oechsle, Michael Niemeyer, Sebastian Nowozin, and
  Andreas Geiger.
\newblock Occupancy networks: Learning 3d reconstruction in function space.
\newblock In {\em Proceedings of the IEEE/CVF conference on computer vision and
  pattern recognition}, pages 4460--4470, 2019.

\bibitem{mildenhall2021nerf}
Ben Mildenhall, Pratul~P Srinivasan, Matthew Tancik, Jonathan~T Barron, Ravi
  Ramamoorthi, and Ren Ng.
\newblock Nerf: Representing scenes as neural radiance fields for view
  synthesis.
\newblock {\em Communications of the ACM}, 65(1):99--106, 2021.

\bibitem{mohan2015timbir}
K~Aditya Mohan, SV Venkatakrishnan, John~W Gibbs, Emine~Begum Gulsoy, Xianghui
  Xiao, Marc De~Graef, Peter~W Voorhees, and Charles~A Bouman.
\newblock Timbir: A method for time-space reconstruction from interlaced views.
\newblock {\em IEEE Transactions on Computational Imaging}, 1(2):96--111, 2015.

\bibitem{nakane2018characterisation}
Takanori Nakane, Dari Kimanius, Erik Lindahl, and Sjors~HW Scheres.
\newblock Characterisation of molecular motions in cryo-em single-particle data
  by multi-body refinement in relion.
\newblock {\em elife}, 7:e36861, 2018.

\bibitem{nogales2016development}
Eva Nogales.
\newblock The development of cryo-em into a mainstream structural biology
  technique.
\newblock {\em Nature methods}, 13(1):24--27, 2016.

\bibitem{park2019deepsdf}
Jeong~Joon Park, Peter Florence, Julian Straub, Richard Newcombe, and Steven
  Lovegrove.
\newblock Deepsdf: Learning continuous signed distance functions for shape
  representation.
\newblock In {\em Proceedings of the IEEE/CVF conference on computer vision and
  pattern recognition}, pages 165--174, 2019.

\bibitem{park2021nerfies}
Keunhong Park, Utkarsh Sinha, Jonathan~T Barron, Sofien Bouaziz, Dan~B Goldman,
  Steven~M Seitz, and Ricardo Martin-Brualla.
\newblock Nerfies: Deformable neural radiance fields.
\newblock In {\em Proceedings of the IEEE/CVF International Conference on
  Computer Vision}, pages 5865--5874, 2021.

\bibitem{peng2021animatable}
Sida Peng, Junting Dong, Qianqian Wang, Shangzhan Zhang, Qing Shuai, Xiaowei
  Zhou, and Hujun Bao.
\newblock Animatable neural radiance fields for modeling dynamic human bodies.
\newblock In {\em Proceedings of the IEEE/CVF International Conference on
  Computer Vision}, pages 14314--14323, 2021.

\bibitem{peng2021neural}
Sida Peng, Yuanqing Zhang, Yinghao Xu, Qianqian Wang, Qing Shuai, Hujun Bao,
  and Xiaowei Zhou.
\newblock Neural body: Implicit neural representations with structured latent
  codes for novel view synthesis of dynamic humans.
\newblock In {\em Proceedings of the IEEE/CVF Conference on Computer Vision and
  Pattern Recognition}, pages 9054--9063, 2021.

\bibitem{punjani2017cryosparc}
Ali Punjani, John~L Rubinstein, David~J Fleet, and Marcus~A Brubaker.
\newblock cryosparc: algorithms for rapid unsupervised cryo-em structure
  determination.
\newblock {\em Nature methods}, 14(3):290--296, 2017.

\bibitem{reed2021dynamic}
Albert~W Reed, Hyojin Kim, Rushil Anirudh, K~Aditya Mohan, Kyle Champley, Jingu
  Kang, and Suren Jayasuriya.
\newblock Dynamic ct reconstruction from limited views with implicit neural
  representations and parametric motion fields.
\newblock In {\em Proceedings of the IEEE/CVF International Conference on
  Computer Vision}, pages 2258--2268, 2021.

\bibitem{rohkohl2008c}
Christopher Rohkohl, Gunter Lauritsch, Alois Nottling, Marcus Prummer, and
  Joachim Hornegger.
\newblock C-arm ct: Reconstruction of dynamic high contrast objects applied to
  the coronary sinus.
\newblock In {\em 2008 IEEE Nuclear Science Symposium Conference Record}, pages
  5113--5120. IEEE, 2008.

\bibitem{scheres2012bayesian}
Sjors~HW Scheres.
\newblock A bayesian view on cryo-em structure determination.
\newblock {\em Journal of molecular biology}, 415(2):406--418, 2012.

\bibitem{schwemmer2013residual}
Chris Schwemmer, Christopher Rohkohl, G{\"u}nter Lauritsch, Kerstin M{\"u}ller,
  and Joachim Hornegger.
\newblock Residual motion compensation in ecg-gated interventional cardiac
  vasculature reconstruction.
\newblock {\em Physics in Medicine \& Biology}, 58(11):3717, 2013.

\bibitem{schwemmer2013opening}
Chris Schwemmer, Christopher Rohkohl, G{\"u}nter Lauritsch, Kerstin M{\"u}ller,
  Joachim Hornegger, and J Qi.
\newblock Opening windows-increasing window size in motion-compensated
  ecg-gated cardiac vasculature reconstruction.
\newblock In {\em Proc. Int. Meeting Fully Three-Dimensional Image
  Reconstruction Radiol. Nucl. Med}, pages 50--53, 2013.

\bibitem{sitzmann2020implicit}
Vincent Sitzmann, Julien Martel, Alexander Bergman, David Lindell, and Gordon
  Wetzstein.
\newblock Implicit neural representations with periodic activation functions.
\newblock {\em Advances in Neural Information Processing Systems},
  33:7462--7473, 2020.

\bibitem{takeishi2017learning}
Naoya Takeishi, Yoshinobu Kawahara, and Takehisa Yairi.
\newblock Learning koopman invariant subspaces for dynamic mode decomposition.
\newblock {\em Advances in neural information processing systems}, 30, 2017.

\bibitem{tretschk2021non}
Edgar Tretschk, Ayush Tewari, Vladislav Golyanik, Michael Zollh{\"o}fer,
  Christoph Lassner, and Christian Theobalt.
\newblock Non-rigid neural radiance fields: Reconstruction and novel view
  synthesis of a dynamic scene from monocular video.
\newblock In {\em Proceedings of the IEEE/CVF International Conference on
  Computer Vision}, pages 12959--12970, 2021.

\bibitem{van2005fourier}
Marin Van~Heel and Michael Schatz.
\newblock Fourier shell correlation threshold criteria.
\newblock {\em Journal of structural biology}, 151(3):250--262, 2005.

\bibitem{wang2021imaging}
Huan Wang, Hima~Manasa Kandula, Ye-Jin Kim, Oh-Hoon Kwon, and Steve Granick.
\newblock Imaging individual molecules using liquid-phase tem-surprises and
  research opportunities.
\newblock {\em Microscopy and Microanalysis}, 27(S2):3--4, 2021.

\bibitem{wang2022neural}
Yuehao Wang, Yonghao Long, Siu~Hin Fan, and Qi Dou.
\newblock Neural rendering for stereo 3d reconstruction of deformable tissues
  in robotic surgery.
\newblock In {\em Medical Image Computing and Computer Assisted
  Intervention--MICCAI 2022: 25th International Conference, Singapore,
  September 18--22, 2022, Proceedings, Part VII}, pages 431--441. Springer,
  2022.

\bibitem{wu2020liquid}
Hanglong Wu, Heiner Friedrich, Joseph~P Patterson, Nico~AJM Sommerdijk, and
  Niels De~Jonge.
\newblock Liquid-phase electron microscopy for soft matter science and biology.
\newblock {\em Advanced materials}, 32(25):2001582, 2020.

\bibitem{yingzhen2018disentangled}
Li Yingzhen and Stephan Mandt.
\newblock Disentangled sequential autoencoder.
\newblock In {\em International Conference on Machine Learning}, pages
  5670--5679. PMLR, 2018.

\bibitem{zang2018space}
Guangming Zang, Ramzi Idoughi, Ran Tao, Gilles Lubineau, Peter Wonka, and
  Wolfgang Heidrich.
\newblock Space-time tomography for continuously deforming objects.
\newblock {\em ACM Trans. Graph.}, 37:1--14, 2018.

\bibitem{zang2019warp}
Guangming Zang, Ramzi Idoughi, Ran Tao, Gilles Lubineau, Peter Wonka, and
  Wolfgang Heidrich.
\newblock Warp-and-project tomography for rapidly deforming objects.
\newblock {\em ACM Transactions on Graphics (TOG)}, 38(4):1--13, 2019.

\bibitem{zhong2021cryodrgnnm}
Ellen~D Zhong, Tristan Bepler, Bonnie Berger, and Joseph~H Davis.
\newblock Cryodrgn: reconstruction of heterogeneous cryo-em structures using
  neural networks.
\newblock {\em Nature methods}, 18(2):176--185, 2021.

\bibitem{zhong2019reconstructing}
Ellen~D Zhong, Tristan Bepler, Joseph~H Davis, and Bonnie Berger.
\newblock Reconstructing continuous distributions of 3d protein structure from
  cryo-em images.
\newblock {\em arXiv preprint arXiv:1909.05215}, 2019.

\bibitem{zhong2021cryodrgn2}
Ellen~D Zhong, Adam Lerer, Joseph~H Davis, and Bonnie Berger.
\newblock Cryodrgn2: Ab initio neural reconstruction of 3d protein structures
  from real cryo-em images.
\newblock In {\em Proceedings of the IEEE/CVF International Conference on
  Computer Vision}, pages 4066--4075, 2021.

\end{thebibliography}
}

\end{document}